\documentclass[showpacs,superscriptaddress,floatfix,amsmath,amssymb,twocolumn,pre]{revtex4-1}
\usepackage{epsfig} 
\usepackage{graphicx}
\usepackage{dcolumn}
\usepackage{bm}
\usepackage[normalem]{ulem} 
\usepackage{amsmath}
\usepackage{amssymb}
\usepackage{epstopdf}
\usepackage{caption}
\usepackage{subcaption}
\usepackage[caption=false]{subfig}
\usepackage{color}

\newcommand{\qb}{{\bf q}}

\newcommand{\qbS}{{\bf q}_{\cal S}}
\newcommand{\qS}{q_{\cal S}}
\newcommand{\qbA}{{\bf q}_A}

\newcommand{\fb}{{\bf f}}
\newcommand{\Kb}{{\bf K}}
\newcommand{\cb}{{\bm \chi}}

\usepackage{color}
\usepackage{ulem}
\def\<{\lesssim}
\def\>{\gtrsim}

\begin{document}

\title{Noise Induced Switching and Extinction in Systems with  Delay}

\author{Ira B. Schwartz}
\affiliation{US Naval Research Laboratory,Code 6792,Nonlinear System Dynamics Section,Plasma Physics Division,Washington, DC 20375}
\author{ Lora Billings}
\affiliation{Department of Mathematical Sciences, Montclair State University, Montclair, NJ 07043}
\author{ Thomas W. Carr}
\affiliation{Department of Mathematics, Southern Methodist University, Dallas, TX 75275}
\author{M. I. Dykman}
\affiliation{Department of Physics and Astronomy,Michigan State University,East Lansing, MI 48824
}

\begin{abstract}
We consider the rates of noise-induced switching between the stable states of dissipative
dynamical systems with delay and also the rates of noise-induced extinction, where such systems model population dynamics. We study a class
of systems  where the evolution depends on the  dynamical
variables at a preceding time with a fixed time delay, which we call hard delay.  For weak noise, the rates
of inter-attractor switching and extinction are exponentially small. Finding these rates to logarithmic accuracy is reduced to variational problems.  The solutions of the variational
problems  give the most probable paths
followed in switching or extinction.  We
show that the equations for the most probable paths are acausal and formulate
the appropriate boundary conditions. Explicit general results are obtained for small delay compared to the relaxation rate.
We also develop a direct variational method to find the rates. 
We find that the analytical results agree well with the
numerical simulations for both switching and extinction rates. 

\end{abstract}

\maketitle


\section{Introduction}

Many physical systems of current interest, including population systems, are
characterized by delay. The evolution of such systems depends not only on the
values of their dynamical variables at a given time, but also on the values
these variables took at previous times. Often
there is a single time delay fixed to a
certain value, which we call a hard-delay case. Well-known examples are
provided by optical systems such as  ring lasers or lasers with external cavities, where the delay is determined by the duration of light propagation along a
certain path, 
cf. \cite{Ikeda1989,Arecchi1991,Heil2001,*Heil2003,Ray2006,Franz2007,Uchida2008,Soriano2013,Marandi2014}. Another
example is Josephson junctions coupled to shunting transmission lines
\cite{Esteve1986,Grabert1988a}.  Examples in biology include various systems,
such as   neural
networks and genetic networks with delay \cite{Marcus1989,Gupta2014}, and
others.  In population  models, \cite{Cushing1977,
  Taylor2009,Blyuss2010,Wang2014},  the delay can be related to temporary
immunity in disease propagation or the time between conception and birth.  Evidence of the pervasiveness of delays across science disciplines
can be seen in several reviews, e.g., \cite{TE2009}. A qualitative stability analysis of delayed systems
can be found in Refs.~\onlinecite{BC1963,AK1963,JH1977}.

An important role in the systems mentioned above is played by noise. Delayed
dissipation and fluctuations of systems coupled to a thermal bath have been
discussed starting from the mid-60s \cite{Mori1965,Kubo1966}, and several
interesting delay-related features of fluctuations have been found, see
\cite{Grote1980,Pollak1986,Reimann2001,Dhar2007,Maes2013} and references
therein. This work focused primarily on the delay described by a retarded
friction force of the form of an integral of the velocity over the time
preceding the observation time; such forces naturally arise in systems
linearly coupled to a thermal bath of harmonic oscillators as well as from
hydrodynamic memory effects
\cite{Franosch2011,Kheifets2014,Donev2014}. 
Systems with hard delay have different physics behind them. In the analysis of
such systems,  much attention has been paid to describing the system
behavior by an effective Fokker-Planck equation,
cf. \cite{Guillouzic1999,Guillouzic2000,Frank2002,Frank2005}. 

A complete description of
fluctuations in the presence of delay is complicated by the fact that the
system phase space is infinite dimensional. Even without noise, to
predict the future one
has to know not just the instantaneous values of the dynamical variables, but
their values in the past on a finite or possibly infinite interval of time. This limits the
applicability of the description of the dynamics in terms of the probability
distribution in a finite-dimensional space.

We will be primarily interested in  fluctuations in dynamical systems with 
hard delay that have one or more simple stationary states. Without noise such
systems asymptotically approach one of the stable states depending
  on initial history.  If the noise is
weak, on average it causes small fluctuations about the occupied stable
state. However, occasionally there occur large outbursts of noise that drive
the system far from the stable state in the space of dynamical variables or
can result in switching from the occupied stable state to another stable
state.

An important qualitative outcome of large noise outbursts, which plays a
critical role in understanding population dynamics and more generally,
reaction systems, is extinction of one of the species. To the best of our
knowledge, there is no analytical theory of weak-noise induced extinction in the presence of delay.
When a (generally, multi-species) population is
described by a dynamical system, extinction corresponds to reaching a
stationary state on the boundary of an attraction basin where one of the dynamical variables is zero;
this variable describes the population of the species that goes extinct. In
our analysis, we will use the term ``extinction" in this meaning.

The idea of our approach goes back to Feynman \cite{FeynmanQM}, who noticed
that, in noise driven systems, even though the noise is random, each noise
realization leads to a certain system trajectory. Therefore, the probability
density of realizations of system trajectories is determined by the
probability density of realizations of the noise trajectories. To find the rate of occurrence of a rare event one has to look for the most
probable realization of the noise trajectories that bring the system to the desired state.  

In what follows we formulate the variational problem for the most probable
path followed by the system on its way to a desired state. Its solution gives
the exponents of the rates of rare events. The possibility to find the most
probable path in the presence of a delay is rooted in the fact that, prior to
the rare event, the system spends a long time performing small fluctuations
about the occupied stable state. This time is much longer than the delay, if
the noise is weak on average. This alleviates the problem of initial
conditions  for trajectories in systems with delay. Our formulation includes
the analysis of switching between the stable states as well as extinction of a
species. We show that these two cases should be analyzed differently.

The significant effect of delay on extinction is physically very
clear. Indeed if, as a result of a fluctuation, the species has gone
extinct at a given instant of time, prior to this time the species population
was nonzero. Therefore the population can come back into existence. The idea can be illustrated by a simple example from the delayed population growth problem: if you sacrifice all of the chickens but keep the eggs, you will
still have chickens after the eggs hatch. 

Previously we discussed the rates of rare events for a one-dimensional
particle with inertia and a retarded friction force described by an integral
of the velocity over time with a weighting factor \cite{Dykman2012a}. We formulated a variational problem for the most probable path followed in switching between stable states in this case. In Ref.~\cite{Lindley2013} we obtained a numerical solution for the most probable switching path in a model system with hard delay and compared the results with Monte-Carlo simulations, with adjusted parameters.

In contrast to the numerical work on a particular system with hard delay, here we are interested in the general features of the effect of hard delay on the rates of rare events. This includes the possibility of developing a perturbation theory in the delay. It turns out  that, because of the difference in the character of the delay, the predictions of the perturbation theory we develop here are qualitatively different from those in systems studied earlier \cite{Dykman2012a}. We also consider the problem of the extinction rate in the presence of delay, which requires a significant extension of both the analytical theory and the Monte Carlo simulation technique, as it is necessary to consider large rare fluctuations induced by a singular multiplicative noise.  The other range of questions we are addressing here concerns the onset of scaling behavior of the switching and extinction rate. Yet another general question is whether delay increases or decreases the rates of switching and extinction.

The layout of the paper is as follows. In Sec.~\ref{section:general_formulation}, we formulate a dynamical model that
describes fluctuations in systems with hard delay. We discuss the stability of
stationary states in the presence of delay and consider small-amplitude
fluctuations about the stable states  due to weak on average noise. In
Sec.~\ref{section:Variation} we formulate a variational approach to the
problem of reaching a state in the space of dynamical variables, which is
remote from the initially occupied stable state. We extend this formulation to
consider the rates of switching between stable states and the problem of
extinction. In Sec.~\ref{sec:special} we find explicit solutions for the rates of
switching and extinction near bifurcation points, including the critical
exponents that describe the scaling of the rates with the distance to the
bifurcation point. We develop a perturbation theory in the ratio of the delay to the relaxation time of the system. 
We also develop a direct variational method for the analysis of the rates. In Sec.~\ref{sec:additive} we compare our theory of the switching rates with
detailed parameter-free numerical simulations, whereas in Sec.~\ref{sec:multiplicative} we
compare the theory and numerical simulations for the problem of extinction. We
finish the paper with a discussion in Sec.~\ref{sec:discuss}.

\section{Model of a delayed noise-driven system}
\label{section:general_formulation}

We consider a system with dynamical variables $\qb=(q_1, q_2,\ldots,q_N)$ and with delay time $\tau$. Its stochastic dynamics are described by the Langevin equation
\begin{equation}
\label{eq:DDE_1}
\dot{\qb}(t)=\Kb\bigl(\qb(t),\qb(t- \tau)\bigr)+ \hat G(\qb(t)) \fb (t).
\end{equation}
Here, $\Kb: {\mathbb{R}}^N \times {\mathbb{R}}^N \rightarrow {\mathbb{R}}^N$ defines the noise-free evolution, whereas
$\fb(t)\equiv \left(f_1(t),f_2(t),\ldots.f_M(t)\right)$ is the $M$-dimensional
noise vector with zero-mean components $f_m(t)$, and $ \hat{G}(\qb(t))$ is an $N \times M$ matrix such that $M \leq N$.  We assume the noise to be Gaussian, stationary, and weak on average. For brevity, we use notations $\qb_\tau \equiv \qb(t-\tau)$ and $\qb_{-\tau} \equiv\qb(t+\tau)$.

We assume that the noise-free equation with delay  $\dot\qb = \Kb(\qb,\qb_\tau)$ has a stationary stable state (attractor) $\qb_A$, near where the system is initially located, and a saddle point $\qbS$. The stationary states satisfy $\Kb(\qb_A,\qb_A) = \Kb(\qbS,\qbS)={\bm 0}$. 

The stability of each stationary state is given by the linearized equations of motion about that state. They have the form
\begin{align}
\label{eq:linearization}
&\dot{{\bm X}}(t)=\hat{\cal K}^{(1)}{\bf X}(t)+ \hat{\cal K}^{(2)}{\bf X}(t-\tau),\nonumber\\
&{\cal{K}}^{(1)}_{nn'}=\partial K_n/\partial q_{n'}, \qquad  {\cal K}^{(2)}_{nn'}=\partial K_n/\partial (q_\tau)_{n'}.
\end{align}
The partial derivatives of ${\bf K}(\qb, \qb_\tau)$ that determine the matrices $\hat{\cal K}^{(1)}$ and $\hat{\cal K}^{(2)}$ are evaluated at ${\qb}=\qb_\tau =\qb_A$ or $\qb=\qb_\tau=\qb_S$. Seeking solution of Eq.~(\ref{eq:linearization}) in the form ${\bf X}(t)=e^{\alpha t}{\bf v}$ leads one to the eigenvalue problem 
\begin{eqnarray}
\label{eq:eigenvalues_stable}
h(\alpha, \tau)=\det\left[\alpha \hat{\cal{I}} -\hat{\cal K}^{(1)} - \exp(-\alpha \tau)\hat{\cal K}^{(2)}\right] =0,
\end{eqnarray} 
where $\cal I$ is the identity matrix.

\subsection{Stability in the case of small delay} 
\label{subsec:stability}

From here on, we assume that in the zero delay case, $\tau=0$, the attractor
has all eigenvalues $\alpha_i$ with negative real part. The saddle has only
one positive real eigenvalue (associated with an unstable direction in the
space of dynamical variables), while the rest of the eigenvalue spectrum lies
in the left hand side of the complex plane. In general, delays have the
tendency to destabilize existing attractors. We will assume that this does not
happen, which is the case when delays are small. Stability of the small delay
case can be discerned directly from Eq.~\eqref{eq:eigenvalues_stable}. The
result for small delay is proven rigorously in Bellman and Cook
\cite{BC1963}. It is shown in Ref.~\onlinecite{BC1963} that there is a finite
range of the values of $\tau$ such that in this range the roots of the
characteristic equation $h(\alpha,\tau)=0$ are either close to the roots of
$h(\alpha,0)=0$ or that the perturbed roots have more negative real parts than
those satisfying $h(\alpha,0)=0$. This indicates that small delays do not
change the asymptotic stability of the attractors in the absence of delays.

It was also proven in Ref.~\onlinecite{BC1963} that simple roots of
$h(\alpha,\tau)$  change continuously  with $\tau$ given that $h(\alpha,\tau)$
is analytic in $\alpha$ in a certain finite range around the corresponding
roots for $\tau = 0$ and $\partial_\tau h(\alpha,\tau)$ is continuous. This
means that, if motion near the saddle is characterized by one positive root
for $\tau=0$, it will still have one positive root in a certain range of small
$\tau$.

We will assume that, for small $\tau$, no new asymptotically stable states emerge compared to the case $\tau=0$. Of primary interest to us will be two cases. One is the case where the system has two attractors along with the saddle point $\qbS$.  The basins of attraction can be built in a standard way by moving away from attractors in negative time (while holding the system near the attractor for time $\tau$). These basins are separated by the separating manifold which contains $\qbS$. For $\tau=0$ the system moves away from $\qbS$ in positive time along the eigenvector ${\bf e}_u$ that corresponds to the positive eigenvalue $\alpha$.  For $\qb$ close to $\qbS$, depending on the sign of $(\qb - \qbS){\bf e}_u$ the system will approach one or the other attractor. This is also true for small $\tau$ given that the system stays near $\qbS$ for time $\tau$, since the system is characterized by only one positive eigenvalue near $\qbS$; the trajectories from $\qbS$ to the attractors will just slightly change.

The other case is where the system has one attractor and a saddle point. However, it is constrained to stay in the space of the dynamical variables on the one side of a hyperplane that goes through the saddle point normal to ${\bf e}_u$. One can think of one of the dynamical variables (which coincides with $(\qb - \qbS){\bf e}_u$ for small $|\qb -\qbS|$) as describing the population of a certain species in a continuous limit. By construction, it may not be negative. 

\subsection{Small-amplitude fluctuations about the stable states} 
\label{subsec:small_fluctuations}

The stationary zero-mean Gaussian noise $\fb(t)$ is characterized by its time correlation functions $\phi_{nm}(t)$ and their Fourier transforms $\Phi_{nm}(\omega)$, 
\begin{eqnarray}
\label{eq:power_spectrum}
&&\Phi_{nm}(\omega) =\frac{1}{2} \int_{-\infty}^{\infty} dt \exp(i\omega t)\phi_{nm}(t), \\
&&\phi_{nm}(t)= \langle f_n(t) f_m(0)\rangle. \nonumber
\end{eqnarray}
Functions $\Phi_{nm}$ determine the power spectrum of the noise and can be often directly accessed in an experiment. The noise intensity $D$ can be defined as twice the maximal eigenvalue of the matrix $\hat\Phi(\omega)$ with matrix elements $\Phi_{nm}$. Since the noise $\fb(t)$ is stationary, matrix $\hat\phi(t)$ with matrix elements $\phi_{nm}(t)$ has an obvious property $\hat\phi^\dagger (t) = \hat\phi(-t)$; in fact, matrix $\hat\phi(t)$ has real matrix elements, and the Hermitian conjugate matrix $\hat\phi^\dagger$ is just the transposed $\hat \phi$.

For small noise intensity, over a characteristic relaxation time the system  approaches an attractor and then mostly performs small-amplitude fluctuations about it. For a given attractor $\qbA$ these fluctuations can be described by linearizing equation of motion (\ref{eq:DDE_1}) about $\qbA$ and assuming that $q(t)\to \qbA$ for $t\to -\infty$. Changing to Fourier components $\delta\qb(\omega)=\int_{-\infty}^\infty  dt\exp(i\omega t)[\qb(t)-\qbA]$ and $\fb(\omega)=\int_{-\infty}^\infty dt\exp(i\omega t)\fb(t)$, we find 
\begin{eqnarray}
\label{eq:Fourier_interrelation}
&&\delta\qb(\omega)=\hat{\cal G}(\omega)\fb(\omega),\\
&&\hat{\cal G}(\omega)=-\left(i\omega \hat{\cal I} + \hat {\cal K}^{(1)} + \hat{\cal
 K}^{(2)}e^{i\omega \tau} \right)^{-1} {\hat G}(\qbA),\nonumber 
\end{eqnarray}
where the matrix elements of matrices $\hat {\cal K}^{(1,2)}$ are given by Eq.~(\ref{eq:eigenvalues_stable}). 

From Eq.~(\ref{eq:Fourier_interrelation}), the correlator of small-amplitude fluctuations of the delayed system is
\begin{equation}
\label{eq:Gauss_small_fluct}
\langle \delta q_n(\omega)\delta q_m(\omega')\rangle = 4\pi [{\cal G}(\omega)\hat \Phi(\omega){\cal G}^{\dagger}(\omega)]_{nm}\delta (\omega +\omega'). 
\end{equation}
Equations (\ref{eq:Fourier_interrelation}) and (\ref{eq:Gauss_small_fluct}) fully describe small noise-induced fluctuations about the attractor in the presence of delay. The fluctuations are Gaussian, as in the case where there is no delay. Their probability distribution is determined by the matrix with matrix elements (\ref{eq:Gauss_small_fluct}). This distribution and the power spectrum of the fluctuations explicitly depend on the delay time $\tau$. 
 
\section{Variational Problem for Large Rare Fluctuations}
\label{section:Variation}
Even though the noise is weak on average, occasionally there occur large
outbursts of the noise that can drive the system far away from the attractor
and also lead to switching between coexisting attractors. The analysis of the
probability distribution far away from an attractor and of the switching rates
in systems with hard delay and multiplicative noise can be done by extending
the analysis for systems driven by additive noise in the absence of delay or
with the delay due to linear coupling to a thermal reservoir
\cite{Dykman1990,Dykman2012a}. The theorems about the stationary states in the
presence of small hard delay facilitate this analysis. 

The key ideas are that, first, prior to a large fluctuation, for a long time
the system is performing small-amplitude fluctuations about the initially
occupied attractor. Second, as mentioned in the Introduction, even though the
noise trajectories themselves are random, each noise trajectory leads to a
system trajectory, which is uniquely defined by equation of motion
(\ref{eq:DDE_1}) \cite{FeynmanQM}. Third, since the probability densities of
different noise trajectories are exponentially different, in a fluctuation to
a given state the system most likely follows a well-defined trajectory, which
is called the optimal trajectory and which corresponds to the most probable
appropriate noise realization. For systems without delay optimal trajectories
followed in interstate switching have been observed in experiments
\cite{Hales2000, Ray2006, Chan2008}. In fact the actual trajectories form a
narrow tube centered at the optimal trajectory. This is also true for
trajectories followed in a rare fluctuation to any state far from the
attractor. Here, the probability of realization of such a tube of trajectories
gives the stationary probability density of reaching this state. 

We now apply these ideas to the problem at hand. The probability density of a realization of Gaussian noise $\fb(t)$ is given 
by the  probability density functional ${\cal P}_{\fb}[\fb(t)]= \exp\left(-{\cal R}_{\fb}/D\right)$ \cite{FeynmanQM}, where
\begin{equation}
\label{eq:Gauss_Functional}
{\cal R}_{\fb}[\fb(t)]= \frac{1}{4}\int_{-\infty}^\infty dt\,dt'\,\fb(t)\hat{{\cal F}}(t-t')\fb(t'),
\end{equation}
$\hat{{\cal F}}(t)$ is the inverse of the pair correlator of
$\fb(t)$, \hspace{1cm} $\int dt'\hat{\cal F}(t-t')\hat\phi (t'-t'')=2D\hat{\cal
  I}\delta(t-t'')$, and $D$ is the noise intensity. Clearly, $D$ drops out
from the expression for ${\cal P}_\fb$ in terms of $\hat\phi(t)$; however, it
is convenient to have it written explicitly for bookkeeping purposes.

Following the arguments given above, to find the optimal trajectory for reaching a state $\qb_0$ we have to solve the variational problem of minimizing ${\cal R}_{\fb}[\fb(t)]$ (and thus maximizing the value of the  probability density functional ${\cal P}_{\fb}[\fb(t)]$) with the constraint that the system is at the attractor, $\qb(t)\to \qbA$  for 
$t\to -\infty$, whereas  it is found at $\qb_0$ for a given $t=t_0$. This means we have to minimize the functional
\begin{eqnarray}
\label{eq:minimizing_functional} 
&&{\cal R}[\qb,\fb,\cb] = {\cal R}_{\fb}[\fb(t)] \nonumber\\
&&+ \int_{-\infty}^\infty dt \cb(t) [\dot{\qb}(t)-\Kb(\qb,\qb_\tau)-\hat G(\qb)\fb(t) ] .
\end{eqnarray}
Here, $\cb(t)$ is the Lagrange multiplier. It relates the trajectory of the system to the noise trajectory. The minimum is taken with respect to trajectories that go from $\qb(t)\to \qbA, \fb(t)\to 0, \cb(t)\to 0$ for $t\to -\infty$ (the optimal noise realization is $\fb \to {\bf 0}$ where the system is at the attractor for $t\to -\infty$), whereas for a given $t_0$ we have $\qb(t_0)=\qb_0$. 

To logarithmic accuracy, the probability density $\rho(\qb_0)$ to be at point $\qb_0$ is given by the value of ${\cal P}$ for the appropriate noise realization, 
\begin{align}
\label{eq:probability_density}
\rho(\qb_0)&= {\rm const}\times \exp[-R(\qb_0)/D],\nonumber\\
R(\qb_0)&=\min {\cal R}[\qb,\fb,\cb],
\end{align}
where the constant weakly depends on the noise intensity $D$; it is assumed that $\exp[-R(\qb_0)/ D]\ll 1$.

\begin{widetext}
The variational equations for the trajectories $\fb(t), \qb(t),\cb(t)$ that minimize functional ${\cal R}$ are obtained in a straightforward way. One of them, $\delta {\cal R}/\delta \cb(t)=0$, is equation of motion (\ref{eq:DDE_1}). The condition of extremum with respect to $\fb(t)$ gives
\begin{align}
\label{eq:f_optimal}
\frac{\delta{\cal R}}{\delta \fb(t)}= \frac{1}{2}\int_{-\infty}^\infty dt' (\hat{\cal F}(t-t')\fb(t') - \hat G^\dagger \bigl(\qb(t)\bigr)\cb(t))=0.
\end{align}
whereas the condition of extremum with respect to $\qb(t)$ gives 
\begin{align}
\label{eq:Lagrange}
\frac{\delta{\cal R}}{\delta \qb(t)} = -\dot\cb(t) -\partial_{\qb(t)}\Bigl[\cb(t)\cdot \Kb\bigl(\qb(t),\qb(t-\tau)\bigr)+ \cb(t+\tau)\cdot \Kb\bigl(\qb(t+\tau),\qb(t)\bigr)\Bigr]
 - \partial_{\qb(t)}\Bigl(\cb(t) \hat G\bigl(\qb(t)\bigr)\fb(t)\Bigr)=0.
\end{align}
%
Here, differentiation of $\Kb\bigl(\qb(t),\qb(t-\tau)\bigr)$ and $\Kb\bigl(\qb(t+\tau),\qb(t)\bigr)$ over $\qb(t)$ is done keeping $\qb(t\pm\tau)$ constant. Equation (\ref{eq:Lagrange}) is acausal, the value of $\cb(t)$ depends on the values of the dynamical variables on the trajectory at time $t+\tau$. This is a remarkable generic feature of the optimal paths in systems with delay. It is a consequence of the non-locality in time of the variational functional ${\cal R}[\qb,\fb,\cb]$.
\end{widetext}

Since on physical grounds it is clear that the large outburst of noise $\fb(t)$ that causes a large rare fluctuation should decay for $t \to \pm\infty$, Eq.~(\ref{eq:f_optimal}) has an obvious solution
\begin{equation}
\label{eq:solution_for_f}
\fb(t)= D^{-1}\int_{-\infty}^\infty dt'\hat\phi(t-t')\hat G^\dagger\bigl(\qb(t')\bigr)\cb(t').
\end{equation}
It is also clear on physical grounds that the fate of the noise-driven system
after it has reached the target $\qb_0$ does not affect the exponent of the
probability distribution, and this exponent is determined by the probability density to reach $\qb_0$ for the first time. This provides the remaining boundary condition for the variational problem (\ref{eq:minimizing_functional}) and (\ref{eq:probability_density}). For $t>t_0$ we ``disconnect" the system from the noise, $\cb(t)=0$ for $t>t_0$. This is the same boundary conditions as in the absence of delay \cite{Dykman1990}.

The functional ${\cal R}[\qb,\fb,\cb]$ can be associated with mechanical action of an auxiliary dynamical noise-free system. The analogy becomes even more clear if $\fb$ is eliminated using Eq.~(\ref{eq:solution_for_f}) and ${\cal R}$ is written in the form
\begin{align}
\label{eq:Hamiltonian}
&{\cal R} = \int dt \cb(t)\dot\qb (t) - \int dt{\cal H}[\qb,\cb], \nonumber\\
& {\cal H}[\qb,\cb]= \cb(t)\Kb[\qb(t),\qb(t-\tau)]\nonumber\\
&+\frac{1}{2D}\int dt'\cb(t)\hat G\bigl(\qb(t)\bigr)\hat\phi(t-t') \hat G^\dagger \bigl(\qb(t')\bigr)\cb(t').
\end{align}
One can interpret ${\cal H}[\qb,\cb]$ as a nonlocal in time Hamiltonian of the auxiliary system; in what follows we call it effective Hamiltonian. Equations (\ref{eq:DDE_1})  and (\ref{eq:Lagrange}) with $\fb$ expressed in terms of $\cb$ become generalized Hamiltonian equations, with $\qb$ and $\cb$ playing the roles of the coordinate and momentum of the auxiliary system.

\subsection{Switching rate}
\label{subsec:switcing_rate}

Of significant interest to us is the switching rate in a system with hard
delay. For switching to occur, after the noise outburst decays, the system has
to be in the basin of attraction to the other attractor or on its
boundary. Since the noise decays for $t\to\infty$, the system should approach
a stationary state, which is either the attractor or the saddle
point. From Eq.~(\ref{eq:solution_for_f}), the
decay of the noise for $t\to\infty$ requires that either the Lagrange multiplier
$\cb(t)$ also decays for $t\to \infty$,  or
that $\hat G^\dagger \bigl(\qb(t)\bigr) \to 0$, i.e., that $\hat G^\dagger \bigl(\qb(t)\bigr) = 0$ at the corresponding stationary state. This latter case will be
discussed in the next subsection. Here we assume that  $\hat G^\dagger
\bigl(\qb(t)\bigr)$ remains finite. Then near a stationary state for
$t\to\infty$ we have $|\cb(t)|, |\fb(t)|\to 0$. We now linearize
Eq.~(\ref{eq:Lagrange}) near a stationary state,

\begin{align}
\label{eq:chi_linearized}
\dot\cb(t)\approx -\cb(t)\hat{\cal K}^{(1)\dagger} - \cb(t+\tau)\hat{\cal K}^{(2)\dagger},
\end{align}
where matrices $\hat{\cal K}^{(1,2)}$ characterize the motion of the system near a stationary state in the absence of noise and are defined by Eq.~(\ref{eq:linearization}). 

By comparing Eq.~(\ref{eq:chi_linearized}) with the characteristic equation for linearized motion of the system near a stationary state in the absence of noise (\ref{eq:eigenvalues_stable}), we see that the roots of the characteristic equation for $\cb(t)$ are equal to the complex-conjugate roots of the characteristic equation for the noise-free system taken with the opposite sign. As a consequence, by moving along the optimal path the system may not approach an attractor, as $\cb(t)$ will be exponentially increasing there. However, the system may approach the saddle point, where one of the roots of the characteristic equation for noise-free motion is positive, and the corresponding root for Eq.~(\ref{eq:chi_linearized}) is negative. This shows that the endpoint for the optimal trajectory leading to switching is the saddle point. As the system approaches this point, $\cb(t)$ evolves along the vector ${\bf e}_u$ that corresponds to the unstable eigenvector of the linearized noise-free motion. 

To logarithmic accuracy, the rate of switching from attractor $\qb_{A_j}$ to attractor $\qb_{A_{j'}} $ ($j,j'=1,2$)  is 
\begin{equation}
\label{eq:switching_rate}
W_{jj'}={\rm const}\exp(-R_j/D),
\end{equation}
where $R_j=\min{\cal R}[\qb,\fb,\cb]$. The minimum is calculated for trajectories that start from the $j$th attractor with coordinate $\qb_{A_j}$ for $t\to -\infty$ and approach the saddle point $\qbS$ for $t\to\infty$. One can draw an analogy between expression (\ref{eq:switching_rate}) and the familiar Arrhenius expression for the reaction rate in thermal equilibrium \cite{Arrhenius1889} by associating the noise intensity $D$ with temperature (more precisely, $k_BT$) and $R_j$ with the activation energy. In what follows we therefore call $R_j$ the effective activation energy for switching.

\subsection{Extinction rate}
\label{subsec:extinction}

As mentioned above, an important class of problems described by systems with
delay are problems of population dynamics. For example, in the problem of
infection transfer, the delay can be the latent period after which an infected
person becomes contagious. At the mean-field (fluctuation-free) level dynamics
of large populations is often modeled by equations of the type
(\ref{eq:DDE_1}) without the noise term,
cf. Refs.~\onlinecite{TaCa09,Blyuss2010,Wang2014}.

Incorporating noise is a delicate issue even in systems which are spatially
uniform. The noise can come from external sources, like weather fluctuations,
or from the very fact that populations are discrete and in an elementary event
(birth, death, infection) they change by integral numbers rather than
continuously. The elementary events themselves happen at random, and in the
simplest models are described by rates. For small fluctuations this
description can be mapped onto a stochastic equation of the type of
(\ref{eq:DDE_1}), but large rare fluctuations are not adequately described by
this approach, cf. Ref.~\onlinecite{Doering2005}.

In this paper we will assume that the noise comes from external sources and
will consider the problem of extinction of a certain population group, for
example, disease extinction. As explained above, extinction corresponds to one of the populations
becoming equal to zero and in the absence of noise the extinction state is
typically a saddle point.

Of interest is the situation where, once the population has gone extinct, it
is not re-created by the noise. For example, if the infectious disease has been
extinct, it does not re-emerge due to weather fluctuations. Then, if $q_1$ is the
scaled population that goes extinct, at the extinction point $\qbS$ we have
$q_1=0$ and  $G_{1n}(\qbS)=0$. The variable $q_1$ corresponds to the unstable
direction at $\qbS$ in the mean-field picture. Quite generally, near $\qbS$
the appropriate components of the random force intensity $G_{1n}G_{m1}$ should
be linear in $q_1$; this is also the case in the models where the noise is
used to describe the effect of population discreteness
\cite{Herwaarden1995}. Then $G_{1n}\propto q_1^{1/2}$. This behavior can be
also understood if one thinks that the noises acting on different individuals
are independent, and the resulting noise is a sum of these noises, with
intensity proportional to the population, i.e., to $q_1$.

The rate of extinction is determined by the probability per unit time of reaching the hypersurface $q_1=0$. Once it has been reached, the system stays on this hypersurface, since there is no noise that would drive it away. Even though in the mean-field picture the extinct state $\qbS$ is similar to a saddle point, the vanishing of the components $ G_{1n}(\qb)$ for $q_1=0$ leads to a major difference of the extinction problem from the switching problem. Indeed, as seen from Eq.~(\ref{eq:solution_for_f}), the optimal random force decays to zero for $t\to\infty$ if $q_1\to 0$ even where $\chi_1(t)$ remains finite. And in fact $\chi_1(t)$ does remain finite for $t \to\infty$. 

We first show that the variational equations for $\qb(t),\cb(t),\fb(t)$ have a stationary solution with $\qb\to \qbS, \fb\to {\bf 0}$, but with $|\cb|>0$. Indeed, if $\qb(t)\to \qbS$ and $\fb(t)\to {\bf 0}$ for $t\to\infty$, then $\partial_{q_1}G_{1n}\propto q_1^{-1/2}$ diverges. Therefore one can see using Eq.~(\ref{eq:solution_for_f}) that for $t \to\infty$ Eq.~(\ref{eq:Lagrange}) has a stationary solution $\chi_{n>1}(t)\to 0$, whereas $\chi_1$ is given by equation
\begin{align}
\label{eq:chi_1}
\chi_{1{\cal S}}&= -2D\left({\cal K}^{(1)}_{11} + {\cal K}^{(2)}_{11}\right)\nonumber\\
&\times\left\{\partial_{q_1}\left[\hat G(\qb)\int dt \hat\phi (t) \hat G^\dagger(\qb)\right]_{11}\right\}^{-1}_{\qb\to \qbS}.
\end{align}
This means that, in the extinction problem, the optimal trajectory approaches the extinction hypersurface $q_1=0$ with a nonzero momentum $\chi_1$ in the direction normal to the hypersurface. This is similar to what happens in a model white-noise driven system without delay \cite{Herwaarden1995}. With this boundary condition, the rate of extinction becomes
\begin{equation}
\label{eq:extinction_rate}
W_{\rm e} = {\rm const}\times\exp(-R_{\rm e}/D),
\end{equation}
where $R_{\rm e}=\min{\cal R}[\qb,\fb,\cb]$ and the minimum is calculated for trajectories that start from the stable state $\qb_A$ for $t\to -\infty$ and approach the extinction state $\qbS, \chi_{n>1}=0, \chi_1=\chi_{1{\cal S}}$ for $t\to\infty$. We will call $R_{\rm e}$ the effective activation energy of extinction.

\section{Special cases}
\label{sec:special}

\subsection{White-noise driven systems}
\label{subsec:white_noise}

The general formulation simplifies in the case where the noise $\fb(t)$ is white. Without loss of generality we can set the noise correlation matrix to be a unit matrix, $\hat\phi(t)=2D\hat {\cal I}\delta(t)$; the difference between the noise component amplitudes can be incorporated into matrix $\hat G$ in the equation of motion of the system (\ref{eq:DDE_1}). The relation (\ref{eq:solution_for_f}) between $\fb(t)$ and the Lagrange multiplier $\cb(t)$ becomes local in time, $\fb(t)=2\hat G^\dagger\bigl(\qb(t)\bigr)\cb(t)$. The Hamiltonian formulation of the variational problem for optimal paths is particularly convenient in this case, as  the effective Hamiltonian in Eq.~(\ref{eq:Hamiltonian}) becomes
\begin{align}
\label{eq:Legendre}
{\cal H}[\qb,\cb]&= \left[\hat G^\dagger\bigl(\qb(t)\bigr)\cb(t)\right]^2
+ \cb(t)\Kb\bigl(\qb(t),\qb(t-\tau)\bigr).
\end{align}
Because ${\cal H}[\qb,\cb]$  has delay, the equations of motion for $\qb(t)$ and $\cb(t)$ remain acausal.

In the extinction problem, it is important that the random force vanishes at the extinction state. Therefore we consider the white noise in the Ito sense, $\langle \hat G\bigl(\qb(t)\bigr)\fb(t)\rangle ={\bf  0}$.

\subsection{Perturbation theory}
\label{subsec:small_delay}

Because the equations for optimal paths in systems with delay are acausal,
solving them is more complicated than for systems without delay. 
However, if the delay $\tau$ is small, we can consider its effect by perturbation theory, assuming that the variational trajectories $\qb(t),\cb(t)$, and $\fb(t)$ for $\tau >0$ are close to the corresponding trajectories for $\tau = 0$. The overall effect of the delay on the rates of rare events will not necessarily be small since, in the expressions for the rates, the corrections to the effective action $R$ [see Eqs.~(\ref{eq:probability_density}), (\ref{eq:switching_rate}), and (\ref{eq:extinction_rate})]  are divided by a small factor, the noise intensity.
 
We assume that the nonlinear function $\bm K$ is smooth, so that if the trajectory $\qb(t-\tau)\equiv \qb_\tau$ is close to $\qb(t)$, the change of $\Kb$ is small,  $||\bm K(\qb,\qb)-\bm K(\qb,\qb_\tau)||\lessapprox \varkappa||\qb-\qb_\tau||$. We then write the variational functional as 
\begin{equation}
\label{eq:separation_of_functionals}
 {\cal R}[\qb, \fb, \cb] = {\cal R}^{(0)} [\qb, \fb, \cb] + {\cal R}^{(1)}[\qb, \fb, \cb],
\end{equation}
where
\begin{align}
\label{eq:perturbed_functionals}
{\cal R}^{(0)} [\qb, \fb, \cb] &= \frac{1}{4} \int \fb(t)\hat {\cal F}(t-t')\fb(t')dt dt' \nonumber\\
&+ \int \cb(t) [\dot{\qb}-{\bf K}(\qb,\qb)- \hat{G}(\qb)\fb(t) ] dt,\nonumber \\
{\cal R}^{(1)}[\qb,\fb,\cb] &= \int \cb(t) [{\bf K}(\qb,\qb)-{\bf K}(\qb,\qb_\tau)] dt.
\end{align}

The first-order correction to $\min {\cal R}$ can be found by evaluating the value of ${\cal R}^{(1)}$ along the variational trajectory $\qb^{(0)}(t), \cb^{(0)}(t), \fb^{(0)}(t)$ that minimizes ${\cal R}^{(0)}$. For example, the effective activation energy $R_j$ of switching from the state $j$ is $R_j\approx R_j^{(0)} + R_j^{(1)}$, where $R_j^{(0)}=\min {\cal R}^{(0)}[\qb,\fb,\cb]$ (subscript $j$ indicates the boundary conditions for the extreme trajectory of ${\cal R}^{(0)}$),  and to the first order in the delay 
\begin{align}
\label{eq:correction_R_j}
R_j^{(1)} &\approx {\cal R}^{(1)}[\qb^{(0)},\fb^{(0)},\cb^{(0)}] \nonumber\\ 
&\approx \tau\int dt \cb^{(0)}(t)[(\dot\qb\,\partial_{\qb'})\Kb(\qb,\qb')]_{\qb=\qb'=\qb^{(0)}(t)}.
\end{align}
We note that the first line here  gives the correction to $R_j$ not only in the case of small delay $\tau$, but also where the term that contains $\qb(t-\tau)$ in $\Kb$ has a small coefficient. 
Similar expressions describe corrections $R_{\rm e}^{(1)}$ and $R^{(1)}(\qb)$ to the activation energies of extinction and reaching a state $\qb$ far from the attractor.

Sometimes corrections $R_j^{(1)}$ or $R_{\rm e}^{(1)}$ diverge, which indicates the fragility effect \cite{Marthaler2006,Khasin2009,Khasin2012}: a small (but finite) perturbation leads to an exponentially large change of the rate of the rare event that is independent of the perturbation. Formally, the divergence occurs, because $R^{(1)}$ accumulates as $t\to \infty$, which happens near the saddle point in the problem of switching or near the extinction state in the problem of extinction.

An important and nontrivial observation based on Eq.~(\ref{eq:perturbed_functionals}) is that delay does {\it not} lead to the fragility. Since ${\bf K}(\qb,\qb)-{\bf K}(\qb,\qb_\tau)$ exponentially decays as the system approaches the saddle point in the switching problem or the extinction state in the extinction problem, generally the value of the functional ${\cal R}^{(1)}$ does not diverge for small $\tau >0$. 

\subsection{Direct variational method}
\label{subsec:variational}

An approximate method that may be helpful in finding the effective activation energies is the direct variational method. We will consider for simplicity the case of white noise. We assume that the matrix $\hat G(\qb) \hat G^\dagger(\qb)$ is nondegenerate. By construction, it is symmetric, since the matrix $ \hat G(\qb)$ is real. Then the solution of the variational equation for $\cb(t)$ reads
\begin{equation}
\label{eq:variational_cb_white}
\cb(t) = \frac{1}{2} \left[\hat G\bigl(\qb(t)\bigr)\hat G^\dagger \bigl( \qb(t) \bigr) \right]^{-1} [\dot \qb - \Kb\bigl( \qb(t), \qb(t-\tau)\bigr)].
\end{equation}

With this solution, the variational functional ${\cal R}$ becomes
\begin{align}
\label{eq:functional_q_only}
{\cal R}=& \frac{1}{4}\int dt [\dot \qb - \Kb\bigl( \qb(t), \qb(t-\tau)\bigr)] \left[\hat G\bigl(\qb(t)\bigr)\hat G^\dagger \bigl( \qb(t) \bigr) \right]^{-1}\nonumber\\
&\cdot [\dot \qb - \Kb\bigl( \qb(t), \qb(t-\tau)\bigr)]
\end{align}
This functional could be also obtained directly from the functional ${\cal R}_\fb[\fb]$, Eq.~(\ref{eq:Gauss_Functional}), by substituting $\fb(t)$ in terms of $\qb(t)$ from the equation of motion (\ref{eq:DDE_1}). In the case of the problem of reaching a given state $\qb$ one should integrate in Eq.~(\ref{eq:functional_q_only}) from $t\to -\infty$ to $t=0$, whereas in the problem of switching or extinction the upper limit is $t\to \infty$.

By construction, functional (\ref{eq:functional_q_only}) is
non-negative. Therefore to minimize it one can use a direct variational method
by taking a trial trajectory $\qb(t)$ that goes from $\qb_A$ to the target
state and minimizing ${\cal R}$ with respect to the parameters of the
trajectory. We illustrate this approach below.

\subsection{Vicinity of a bifurcation point}
\label{subsec:bifurcation}

The general analysis significantly simplifies in the important parameter range where the system is close to a bifurcation point. We will consider two types of bifurcation points: a saddle-node bifurcation, where the attractor merges with a saddle point, and a very similar transcritical bifurcation commonly considered in the extinction problem, where the attractor merges with the extinction state. Near a bifurcation point the motion is slowed down, there emerges a soft mode (a slow variable) \cite{Arnold1988,Guckenheimer1987}. Therefore the effect of delay is suppressed, generally speaking. As the system approaches a bifurcation point the delay time $\tau$ becomes increasingly smaller than the relaxation time of the system. Moreover, the noise becomes effectively white, as the noise correlation time is small compared to the relaxation time, too. However, there is a parameter range where the delay time exceeds the relaxation time of the fast variables, and yet has not become too short compared to the relaxation time of the slow variable. It is this range that we will discuss.

Quite generally, after the ``fast" variables have approached their stable state for a given value of the slow variable $q$, the equation of motion for this slow variable has the form
\begin{align}
\label{eq:slow_eom}
&\dot q(t) = K\bigl(q(t),q(t-\tau)\bigr) + G\bigl(q(t)\bigr)f(t), \nonumber\\ 
& K(q,q)=-\eta + q^2, \qquad\langle f(t)f(t')\rangle = 2D\delta(t-t').
\end{align}
Parameter $\eta$ gives the distance to the saddle-node bifurcation point in the absence of delay. For $\eta > 0$ the system has a stable state at  $q_A =- \eta^{1/2}$ and a saddle at $q_{\cal S}=\eta^{1/2}$. For $\eta=0$ these states merge, and for $\eta <0$ there are no stationary states in the range of small $|q|$. Close to the bifurcation point $|\eta|\ll 1$. For a transcritical bifurcation one has $K(q,q)=rq-q^2$, in which case for $r>0$ the system has a stable state $q_A = r$ and a saddle point $q_{\cal S}=0$, whereas for $r<0$ in the range $q\geq 0$ the system has only a stable stationary state $q=0$. By changing $q\to q+r/2, r\to 2\eta^{1/2}, K\to -K$ the equation of motion near a transcritical bifurcation can be brought into the same form as the equation of motion near a saddle-node bifurcation point. Therefore in what follows we use Eq.~(\ref{eq:slow_eom}) to consider both cases. 

In the analysis of the switching problem, where $G(q)$ is generally non-singular, taking into account that the characteristic scale of  $q$ is small, one can set $G(q)=1$. Then quite generally, to the first order in $\tau$ we have the following expressions for the optimal trajectory and the activation energy of switching:
\begin{align}
\label{eq:switching_exponent_bifurcation}
&q^{(0)}_{\rm sw}(t)=\eta^{1/2}\frac{e^{2t\sqrt{\eta}} -1}{e^{2t\sqrt{\eta}} +1},\nonumber\\
&R_{\rm sw}\approx \frac{4}{3}\eta^{3/2}  -\tau \int_{-\sqrt{\eta}}^{\sqrt{\eta}}dq K(q,q)\left[\partial_{q'}K(q,q')\right]_{q'=q} , 
\end{align}
here we used Eq.~(\ref{eq:correction_R_j}) and took into account that $\chi^{(0)}_{\rm sw}(t)= -K\bigl(q^{(0)}_{\rm sw}(t),q^{(0)}_{\rm sw}(t)\bigr)$. Interestingly, depending on the form of the $K$-function, the scaling of the $\tau$-dependent correction with the distance to the bifurcation point $\eta$ can be different from that of the main term. If $K\bigl(q(t),q(t-\tau)\bigr)$ is linear in $q(t-\tau) - q(t)$, the scaling of the correction and the main term is the same, $\eta^{3/2}$; if the expansion of $K$ starts with $[q(t-\tau) - q(t)]^2$,  the correction to $R_{\rm sw}$ dies out with decreasing $\eta$ as $\eta^{5/2}$.

For the extinction problem, as explained above, $G(q)=\left(\eta^{1/2}-q\right)^{1/2}$. In this case we obtain the same trajectory in the absence of delay as in the problem of switching, $q^{(0)}_{\rm e}(t)=q^{(0)}_{\rm sw}(t)$, whereas the parameter $\chi^{(0)}$ for the extinction problem is  $\chi^{(0)}_{\rm e}(t)= -K\bigl(q^{(0)}_{\rm sw}(t),q^{(0)}_{\rm sw}(t)\bigr)/G^2\bigl(q^{(0)}_{\rm e}(t)\bigr)$. Then, with account taken of Eq.~(\ref{eq:correction_R_j}), 
\begin{align}
\label{eq:extinction_exponent_bifurcation}
R_{\rm e}\approx 2\eta  -\tau \int_{-\sqrt{\eta}}^{\sqrt{\eta}}dq \frac{K(q,q)}{\sqrt{\eta}-q}\left[\partial_{q'}K(q,q')\right]_{q'=q} . 
\end{align}
The leading-order term in the activation energy (\ref{eq:extinction_exponent_bifurcation}) scales as $r^2\propto \eta$ with the distance to the transcritical bifurcation point. 

The simple expressions (\ref{eq:switching_exponent_bifurcation}) and (\ref{eq:extinction_exponent_bifurcation}) give the activation energies for the switching and extinction problems in the explicit form, to the lowest order in delay. They are particularly important in terms of applications, since the switching rates near bifurcation points are comparatively large and are easier to access in the experiment.

\section{Switching induced by additive white noise in a model system}
\label{sec:additive}

We now apply the general results to simple models where the delay-induced
corrections to the rates of switching and extinction can be calculated
explicitly and the results can be compared to extensive numerical
simulations. We consider two models. Both are described by one dynamical
variable $q$. This variable is assumed to be driven by white Gaussian
noise. In this section we assume that the noise is additive, $G(q)=1$. The
difference between the models is in the form of the delay in the regular
force, which is equal to   $K_{\rm attr}$  or $K_{\rm  rpl}$,
\begin{align}
\label{eq:simple_K}
&K_{\rm rpl}\bigl(q(t),q(t-\tau)\bigr)= -\gamma q(t) - q^2(t) + q(t-\tau),\nonumber\\
&K_{\rm attr}\bigl(q(t),q(t-\tau)\bigr)= -\gamma q(t-\tau) - q^2(t) + q(t),
\end{align}
with $0<\gamma<1$. For the both forms of the force, in the absence of noise and delay the system has a stable state $q_A=1-\gamma$ and a saddle point $q_{\cal S}=0$. We use subscripts ``rpl" and ``attr" to indicate that, in the corresponding model, delay is in the term that describes, respectively, repulsion and attraction to the saddle point $q=0$.  In the presence of delay, for $K_{\rm attr}$ the characteristic equation (\ref{eq:eigenvalues_stable}) for the saddle point along with a positive root $\alpha >0$ has also a negative root $\alpha\approx \tau^{-1}\log\gamma\tau$ for $\gamma\tau \ll 1$. This negative root does not change the results on the switching rate for small $\gamma\tau$, the optimal path corresponds to the generalized momentum $\chi(t)$ varying along the eigenvector that corresponds to the positive root.

For $\tau=0$ the effective Hamiltonian ${\cal H}[q,\chi]\equiv {\cal H}^{(0)}[q,\chi]$, see Eq.~(\ref{eq:Hamiltonian}), for the both models becomes local in time, 
\begin{equation}
\label{eq:Hamiltonian_0}
{\cal H}^{(0)}[q,\chi]=\chi K(q,q)+\chi^2, \quad K(q,q)=(1-\gamma)q -q^2.
\end{equation}
Since the value of ${\cal H}^{(0)}$ is conserved along the zero-delay variational trajectory $q^{(0)}(t),\chi^{(0)}(t)$, and this trajectory starts for $t\to -\infty$ from $q\to q_A, \chi\to 0$, we have ${\cal H}^{(0)}=0$. Then the zero-delay Hamiltonian trajectory is 
\begin{align}
\label{eq:zero_switch_traj}
&q^{(0)}(t)=q_A/[1+\exp(q_At)],\nonumber\\
&\chi^{(0)}(t)=-q_A^2 \exp(q_At)/[1+\exp(q_A t)]^2.
\end{align}
Clearly, $q^{(0)}(t)\to q_{\cal S},\chi^{(0)}(t)\to 0$ for $t\to \infty$, so that Eq.~(\ref{eq:zero_switch_traj}) indeed describes the most probable path followed in switching.

From Eq.~(\ref{eq:Hamiltonian}), the delay-induced correction to the effective Hamiltonian ${\cal H}^{(1)}[q,\chi]=\chi(t)[K\bigl(q(t,q(t-\tau)\bigr) - K\bigl(q(t),q(t)\bigr) ]$ for the two models becomes, respectively, ${\cal H}^{(1)}_{\rm rpl} = -\chi(t) [q(t)-q(t-\tau)]$ and ${\cal H}^{(1)}_{\rm attr} = \gamma\chi(t)[q(t)-q(t-\tau)]$. The first-order correction to the switching activation energy is given by $-\int {\cal H}^{(1)}[q^{(0)}(t),\chi^{(0)}(t)] dt$, cf. Eq.~(\ref{eq:correction_R_j}). 

From Eqs.~(\ref{eq:perturbed_functionals}) and (\ref{eq:correction_R_j}) we find the zeroth-order activation energy $R_{\rm sw}^{(0)}$ of switching from the stable state $q_A$ and the delay-induced corrections $R_{\rm sw}^{(1)}$ for the two models as
\begin{align}
\label{eq:R_switch_model}
&R_{\rm sw}\approx  R_{\rm sw}^{(0)}+ R_{\rm sw}^{(1)},\qquad   R_{\rm sw}^{(0)}=\frac{1}{6}(1-\gamma)^3, \nonumber\\
&R_{\rm sw, rpl}^{(1)}=  \tau R_{\rm sw}^{(0)}, \qquad R_{\rm sw, attr}^{(1)} = -\gamma \tau R_{\rm sw}^{(0)}.
\end{align}
Equation (\ref{eq:R_switch_model}) shows that, where the delay is in the term that pushes the system away from the saddle point, the delay leads to an increase of the activation energy of switching from the stable state and thus to a decrease in the switching rate. In the opposite case, where the delay is in the term that pushes the system towards the saddle point, the effect is opposite.  This can be understood by noticing that the delay-induced change of $K$ is $\propto q(t-\tau) - q(t)$. The latter quantity is positive when the system moves from $q_A$ to $q_{\cal S}$. The corresponding change of the regular force has to be overcome by the noise as it drives the system to $\qS$. In the model with $K_{\rm rpl}$ a stronger noise is needed than without delay, whereas in the model with $K_{\rm attr}$ the needed noise is weaker.  The corresponding noise realizations are exponentially less/more probable.

For the models (\ref{eq:simple_K}), the general approach to finding a correction to the activation energy to first order in $\tau$ can be checked by expanding the delayed term in Eq.~(\ref{eq:simple_K}) as $q(t-\tau)\approx q(t)-\tau\dot q(t)$. Then the equations of motion for the cases of the delay in the repulsing and attracting terms become, respectively, $(1+ \tau)\dot q(t) = K\bigl(q(t),q(t)\bigr) +f(t)$ and $(1-\gamma \tau)\dot q(t) = K\bigl(q(t),q(t)\bigr) +f(t)$. Dividing these equations by $1+ \tau$ and $1-\gamma\tau$, respectively, we reduce them to the standard Langevin form with no delay and with the rescaled regular force and the noise intensity.  One can easily see that the resulting activation energy has the form (\ref{eq:R_switch_model}).

\begin{figure} \hspace*{-0.1in}
\begin{subfigure}{.25\textwidth}
  \includegraphics[width=\linewidth]{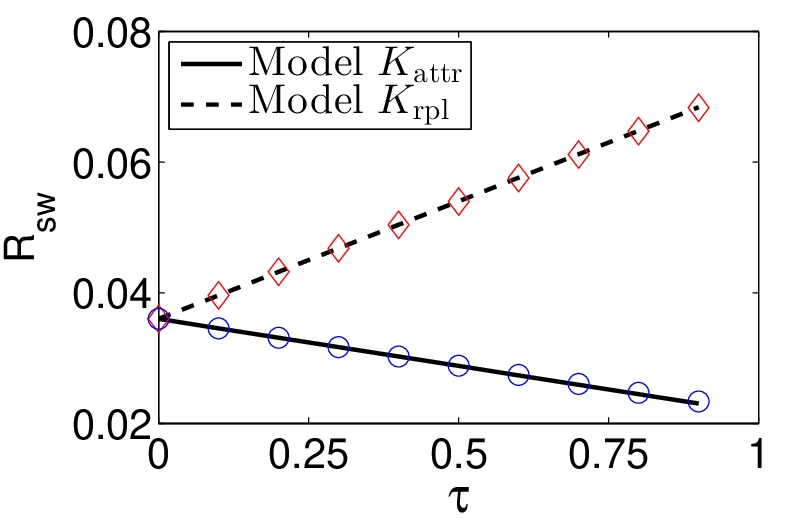}
  \caption{Additive noise}
  \label{fig:Fig1a}
\end{subfigure}%
\begin{subfigure}{.25\textwidth}
  \includegraphics[width=\linewidth]{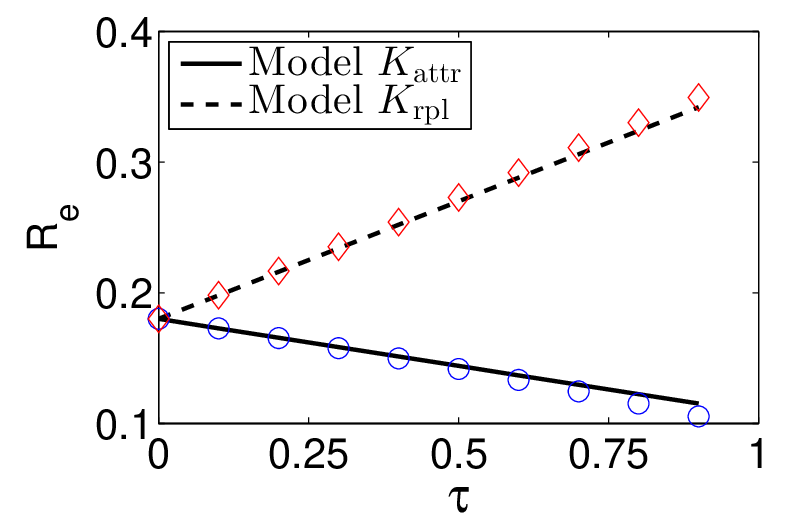}
  \caption{Multiplicative noise}
  \label{fig:Fig1b}
\end{subfigure}
\caption{\small \raggedright A comparison of the activation energies for switching, $R_{\rm sw}$, and extinction, $R_{\rm e}$,  calculated by the
    direct variational method (circles and diamonds) and by the first order
    perturbation theory in the delay $\tau$ (solid and dashed lines) for models  (\ref{eq:simple_K}). Parameter $\gamma$ is fixed at $\gamma=0.4$.}

\label{fig:variational}
\end{figure}

In Fig.~\ref{fig:variational} we compare the $\tau$ dependences of the
activation energy $R$ for the two models, which are calculated by the first
order perturbation theory (\ref{eq:R_switch_model}) and by the direct
variational method. In the variational method, we used a  trajectory of
the form $q_A/[1+\exp(\lambda t)]$ and minimized $R$ with respect to parameter
$\lambda$. The variational results for both switching and extinction activation energies
are in excellent agreement with our perturbation theory in the range of small to moderately small $\tau$, which is of interest for the present paper.

\subsection{Numerical simulations}

In this section, we compare the theoretical results with numerical simulations
for switching.  For the models  (\ref{eq:simple_K}), we looked for the mean
switching time $T_{\rm sw}$ as the white additive noise drives the system out
of the basin of attraction of $q_A$ to $q < q_{\cal S}$ (in our system
$q_A>q_{\cal S}$).  To insure that a trajectory doesn't drift back to the
original basin we require that it goes well past $q_{\cal S}$, so that  the
probability of returning to the basin of attraction is exponentially small. More
specifically, we define the switching to have happened when $q < -0.2$ and
find that this threshold is consistent with the no-return requirement.  We then
calculate $T_{\rm sw}$ as the mean first passage time (MFPT) to $q=-0.2$ given that
the system starts from $q_A$. In each of the
Figs.~\ref{fig:MCtau}, \ref{fig:TswvsD}, and \ref{fig:ExtvsD} below, the
solid and dashed lines represent the first-order in $\tau$ perturbation theory, while the data points are the mean values
taken over 2000 simulations.

\begin{figure}
 \hspace*{-0.1in}
\begin{subfigure}{.25\textwidth}
 
  \includegraphics[width=\linewidth]{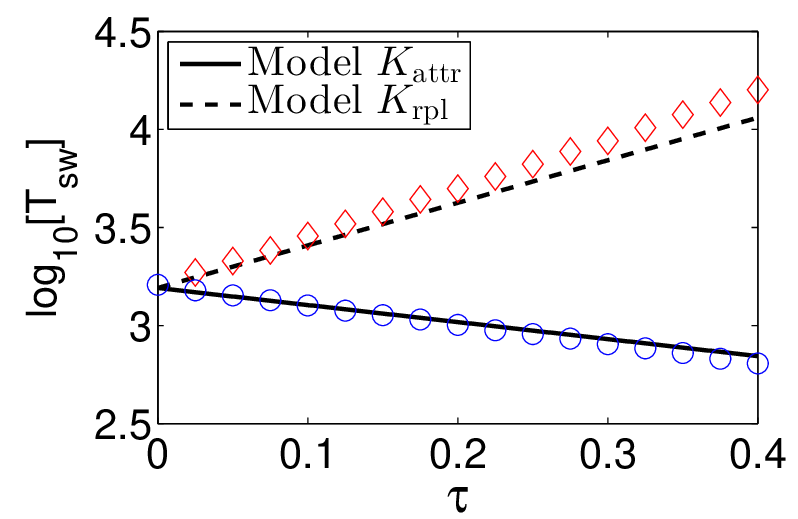}
  \caption{Additive noise with $\sqrt{2D}=0.12$}
  \label{fig:Fig2a}
\end{subfigure}%
\begin{subfigure}{.25\textwidth}
  \centering
  \includegraphics[width=\linewidth]{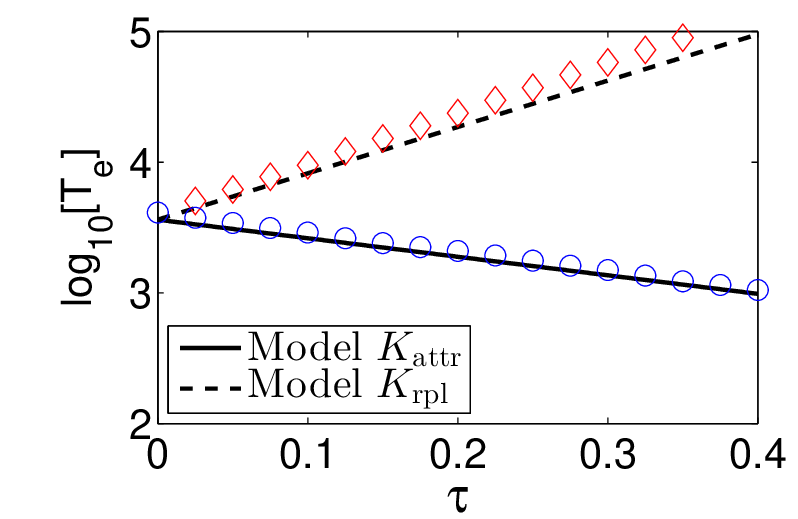}
  \caption{Multiplicative noise with $\sqrt{2D}=0.21$}
  \label{fig:Fig2b}
\end{subfigure}
\caption{\small \raggedright A comparison of the log base 10 of the mean first passage times obtained by Monte Carlo simulations
  (circles and diamonds) and by the
  first order perturbation theory in the delay (solid and dashed lines). The models of the delayed regular force $K_{\rm attr}$ and $K_{\rm rpl}$
are given by Eq.~(\ref{eq:simple_K}), as in the caption of Fig.~\ref{fig:variational},
  and $\gamma=0.4$. }
\label{fig:MCtau}
\end{figure}

Switching events are Poisson-distributed, and the switching time is simply the inverse of the switching rate. Therefore from Eq.~(\ref{eq:switching_rate})
\begin{equation}
\label{eq:T_sw}
T_{\rm sw}=c_{\rm sw}\exp(R_{\rm sw}/D), \qquad c_{\rm sw}\approx 2\pi/(1-\gamma),
\end{equation}
where $R_{\rm sw}$ is the activation energy of switching in
Eq.~(\ref{eq:R_switch_model}).  For the pre-factor $c_{\rm sw}$ we used the
result of the Kramer's theory \cite{Kramers1940} in which there is no delay,
since the major effect of the delay for weak noise is in the logarithm of the MFPT, $R_{\rm sw}/D$.

Similar to the direct variational plots in Fig.~\ref{fig:variational}~(a),
Fig.~\ref{fig:MCtau}~(a) shows the logarithm of $T_{\rm sw}$ as a function of
delay $\tau$. There are no free parameters in the plot. In the model where the regular force is $K_{\rm  attr}$, so that
the delay is in the term that pushes the system toward the saddle point, there is excellent
agreement between the theory and numerical simulations throughout the whole considered range of $\tau$. For the model with $K_{\rm  rpl}$, where delay pushes the system away from the saddle point, 
the theory and simulations are very close for $\tau<0.2$. For larger $\tau$ the discrepancy is somewhat larger, which indicates that, for this model, higher-order delay-induced corrections become important.  Clearly, the delay has opposite effect in the two models: it destabilizes the system in model $K_{\rm  attr}$, leading to the decrease of $T_{\rm sw}$, and stabilizes the system in model $K_{\rm  rpl}$, in agreement with the arguments given above.

A major prediction of the theory is the exponential dependence of the mean switching time on the reciprocal noise intensity $D$. In Fig.~\ref{fig:TswvsD}, we plot the logarithm of the mean switching time as a function of $D^{-1}$ for both models in
Eq.~(\ref{eq:simple_K}) obtained for the delays $\tau=0.25$ and $\tau =0.5$.  Here, too, we have excellent agreement of the
theory and simulations.

\begin{figure}  \hspace*{-0.1in}
\begin{subfigure}{.25\textwidth}
  \includegraphics[width=\linewidth]{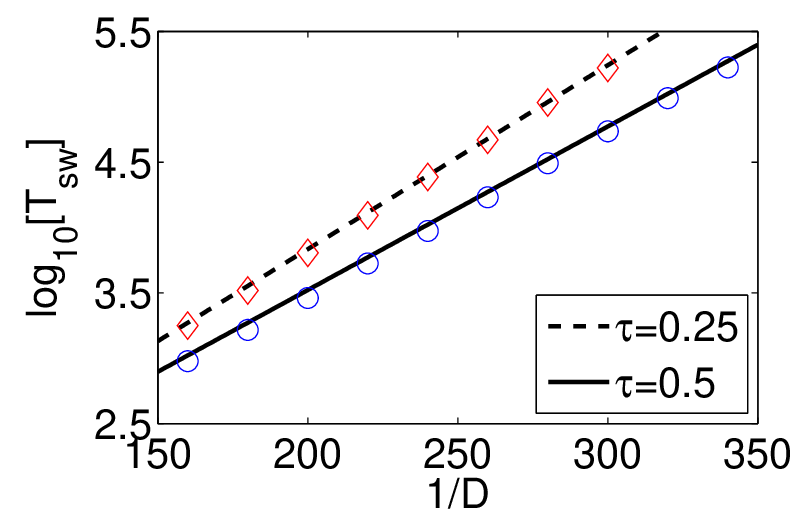}
  \caption{Model $K_{\rm attr}$, additive noise}
  \label{fig:Fig3a}
\end{subfigure}%
\begin{subfigure}{.25\textwidth}
  \includegraphics[width=\linewidth]{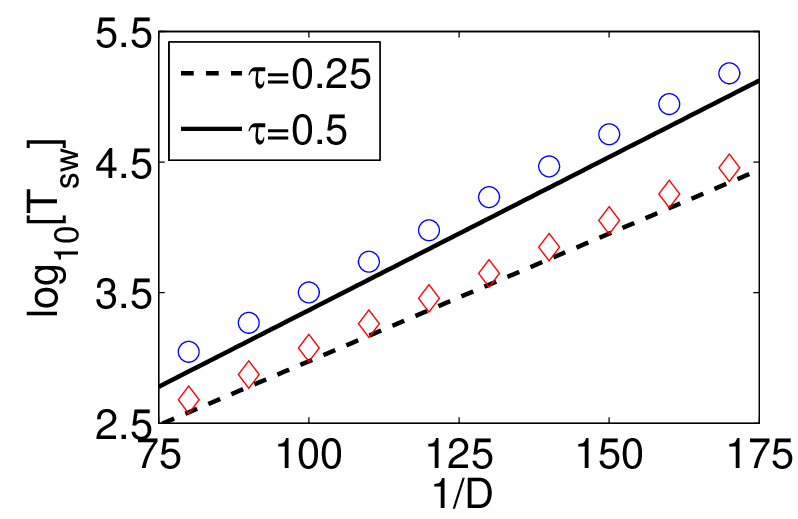}
  \caption{Model $K_{\rm  rpl}$, additive noise}
  \label{fig:Fig3b}
\end{subfigure}
\caption{\small \raggedright Log base 10 of the mean switching time vs. reciprocal noise intensity for the 
 delay  models in Eq.~(\ref{eq:simple_K})  with $\gamma=0.4$. The solid and dashed lines represent the
  theory from Eqs.~(\ref{eq:R_switch_model}) and (\ref{eq:T_sw}), while the data points are the mean values
  taken over 2000 simulations.}

\label{fig:TswvsD}
\end{figure}


\section{Extinction in a model system}
\label{sec:multiplicative}

We now consider the time to extinction. We will use the same models of noise-free motion, Eq.~(\ref{eq:simple_K}).  Without delay and noise, the equation of motion for these models $\dot q = K(q,q)$ with $K=(1-\gamma)q-q^2$, cf. Eq.~(\ref{eq:simple_K}), can be thought of as a logistic equation with linear feedback. Such equation is often used to describe continuous population model. In the population dynamics context, the state $q_{\cal S}=0$ corresponds to the extinction state. As argued in Sec.~\ref{subsec:extinction}, of interest in this case is multiplicative noise. We assume the noise to be $\delta$-correlated and choose the factor $G(q)$ that determines the noise strength in the standard for population dynamics form $G(q)=\sqrt{q}$. 

From Eq.~(\ref{eq:Legendre}), the effective Hamiltonian that describes the most likely trajectory of the system followed in extinction is a sum of the term ${\cal H}^{(0)}$ without delay, which is the same for the both dynamical models and, depending on the model, the terms ${\cal H}_{\rm rpl}^{(1)}$ or ${\cal H}_{\rm attr}^{(1)}$ that come from the delay,
\begin{align}
\label{eq:H_0_extinction}
&{\cal H}^{(0)}[q,\chi] = q\chi^2 + \chi K(q,q), \nonumber\\
&{\cal H}_{\rm rpl}^{(1)}[q,\chi]= \chi(t) [q(t-\tau)- q(t)],\nonumber\\
& {\cal H}_{\rm attr}^{(1)}[q,\chi]= -\gamma\chi(t) [q(t-\tau)- q(t)].
\end{align}
As discussed in Sec.~\ref{subsec:extinction}, and with account taken of Eq.~(\ref{eq:chi_1}), the Hamiltonian trajectory leading to extinction satisfies boundary conditions $q(t)\to q_A = 1-\gamma, \chi(t)\to 0$ for $t\to -\infty$, and $q(t)\to q_{\cal S}, \chi(t)\to -(1-\gamma)$ for $t\to \infty$.

If we disregard delay and use ${\cal H}^{(0)}$ as the Hamiltonian, the Hamiltonian trajectory $q^{(0)}(t)$ is described by Eq.~(\ref{eq:zero_switch_traj}). From condition ${\cal H}^{(0)}=0$ we have $\chi^{(0)}(t)= q^{(0)}(t)-(1-\gamma)$. Using these expressions we obtain from Eqs.~(\ref{eq:correction_R_j}) and (\ref{eq:H_0_extinction}) the activation energy of extinction $R_{\rm e}$ as a sum of the term that describes switching in the absence of delay $R_{\rm e}^{(0)}$ and a correction of the first order in the delay $\tau$. We denote the corrections for the models (\ref{eq:simple_K}) as $R_{\rm e, rpl}^{(1)}$ and $R_{\rm e, attr}^{(1)}$,
\begin{align}
\label{eq:extinct_R}
&R_{\rm e} \approx R_{\rm e}^{(0)}+ R_{\rm e}^{(1)}, \qquad    R_{\rm e}^{(0)}=\frac{1}{2}(1-\gamma)^2, \nonumber\\
& R_{\rm e, rpl}^{(1)}\approx   \tau R_{\rm e}^{(0)}, \qquad R_{\rm e, rpl}^{(1)}\approx  -\gamma\tau R_{\rm e}^{(0)}
\end{align}
This result can be again independently checked by expanding $q(t-\tau)$ in Eq.~(\ref{eq:simple_K}) to the first order in $\tau$, as discussed above.

We also perform a direct variational calculation of $R_{\rm e}$. We used the same variational trajectory as in the switching problem. The results are shown in Fig.~\ref{fig:variational}(b). They are in excellent agreement with the perturbation theory (\ref{eq:extinct_R}) in the studied range of $\tau$.

\subsection{Numerical simulations}

For numerical simulations, we note that our theory of extinction is  based on
the Ito calculus. Therefore, we have used the Milstein method ~\cite{KP1999}, which consistently takes into account that
$\langle G\bigl(q(t)\bigr)f(t)\rangle =0$. To find the mean time of extinction
in the presence of delay, one has to make sure that, after the system has
reached the extinction state in a simulation, it will not leave it. The
possibility to leave is particularly clear for the model (\ref{eq:simple_K}) with $K$ of the form of  $K_{\rm   rpl}$. Indeed, if the system has been brought by the noise to $q=0$ and
after that the noise becomes equal to zero (or just very small), the system
will move away from $q=0$ toward the attractor, because it is driven by the
force $q(t-\tau)>0$.

From the above argument, one cannot use the MFPT to reach $q=0$ as the measure of the reciprocal extinction rate. The system has to stay at $q=0$ for the time equal at least to the delay time. This makes simulations significantly
different from the conventional MFPT simulations.

To make a quantitative comparison of the theory and the simulations we used the prefactor
in the mean time to extinction $T_{\rm   e}$, which was calculated in the
absence of delay.  If there is no delay and the noise is white, $T_{\rm e}$ is
the MFPT for reaching the extinction state from a point $q$ in the vicinity of
the attractor $q_A$. The equation for the MFPT $T(q)$ in our model reads 
$Dq\partial_q^2 T + K(q,q)\partial_qT=-1$, cf. Ref.~\cite{Risken1989}. The solution
is $T(q)=D^{-1}\int_0^q dq_1\int_{q_1}^\infty
dq_2(1/q_2)\exp\{[V(q_1)-V(q_2)/D]\}$ with
$V(q)=\frac{1}{2}q^2-(1-\gamma)q$. For $q$ near $q_A$, the integrand in the
integral over $q_2$ has a maximum near $q_A$ and is Gaussian near the
maximum. The integrand in the integral over $q_1$ has a maximum at $q_1=0$, but it is not Gaussian near the maximum. This is why the result for the prefactor differs from that in the switching problem. Expanding the exponent in this integrand to a linear term in $q_1$
near $q_1=0$, we obtain $T_{\rm e}=c_{\rm e}\exp[R_{\rm e}^{(0)}/D]$
with $c_{\rm e}=(2\pi D)^{1/2}/(1-\gamma)^2$. In plotting the theoretical data we used 
$T_{\rm e}=c_{\rm e}\exp[(R_{\rm e}/D]$ with $R_{\rm e}$ given by Eq.~(\ref{eq:extinct_R}) and the above value of $c_{\rm e}$, which refers to $\tau=0$.  

We first compare the simulations and the perturbation theory of the mean extinction
time $T_e$ as a function of the delay $\tau$. The results are shown in Fig.~\ref{fig:MCtau}~(b). As in the additive noise case, we see
slight disagreement for model $K_{\rm  rpl}$ as the delay gets large, but excellent
agreement for $\tau<0.2$. In contrast, model $K_{\rm  attr}$ shows excellent
agreement for the whole range of $\tau$ we have explored. Again, the theoretical curves have no free parameters. 

\begin{figure} \hspace*{-0.1in}
\centering
\begin{subfigure}{.25\textwidth}
  \centering
  \includegraphics[width=\linewidth]{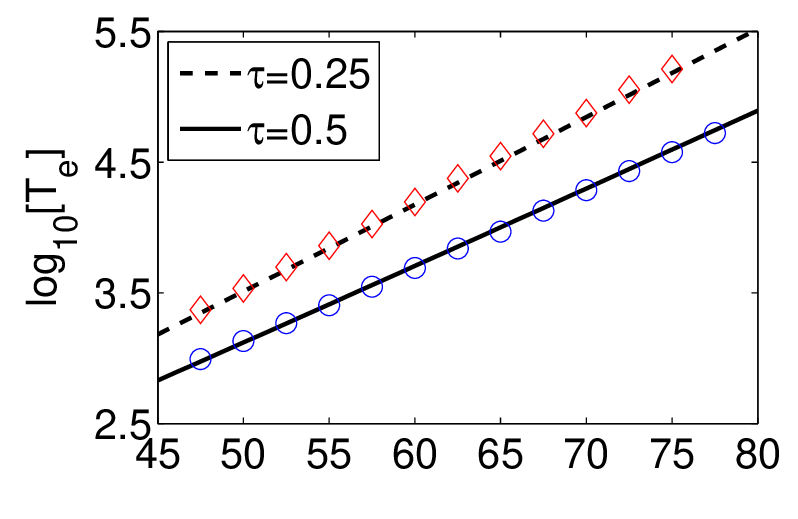}
  \caption{Model  $K_{\rm attr}$, \\ multiplicative noise}
  \label{fig:Fig4a}
\end{subfigure}%
\begin{subfigure}{.25\textwidth}
  \centering
  \includegraphics[width=\linewidth]{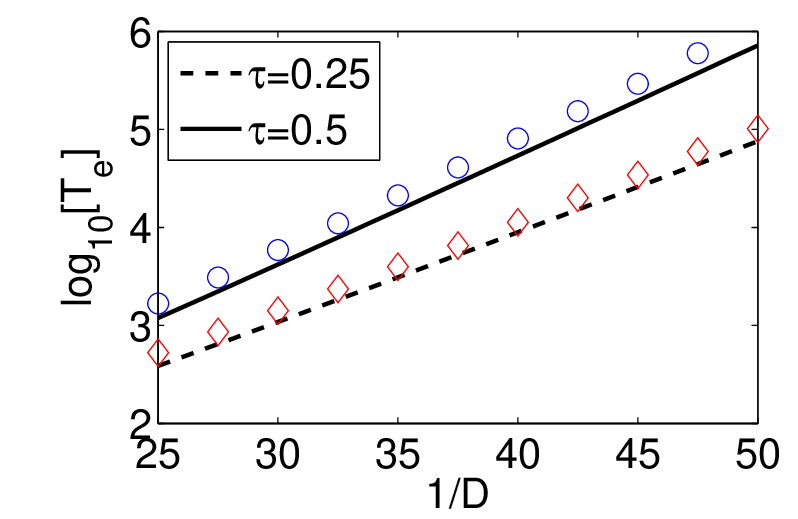}
  \caption{Model $K_{\rm  rpl}$, \\ multiplicative noise}
  \label{fig:Fig4b}
\end{subfigure}
\caption{\small \raggedright Log of extinction time vs. inverse noise intensity for the  models of delayed dynamics in Eq.~(\ref{eq:simple_K}); $\gamma=0.4$. The lines represent the theory, Eq.~(\ref{eq:extinct_R}), while the data points are the mean values  taken over 2000 simulations.}

\label{fig:ExtvsD}
\end{figure}

We also considered the effect of noise intensity on the mean extinction times,
which is depicted in Fig. ~\ref{fig:ExtvsD}. Here again we see the characteristic $1/D$ dependence of $\log T_{\rm e}$, and the results of the theory and the simulations are in excellent agreement for the both models.


\section{Conclusions}
\label{sec:discuss}

We have considered dynamical systems driven by weak on average noise and studied the effect of delay on small fluctuations about the stable states and on the probabilities of large rare fluctuations. Of central interest was the effect of delay on the rates of noise-induced switching between coexisting stable states and noise-induced extinction understood as reaching a saddle point on the boundary of the basin of attraction of a stable stationary state; at this saddle point one of the dynamical variables ($q_1$) is zero; we assumed that, even in the presence of the noise,  once this variable has become equal to zero it remains equal to zero. 

In our analysis delay was incorporated into the effective force that drives the system in the absence of noise. This force depends not only on the instantaneous values of the dynamical variables, but also on the values of these variables a certain time $\tau$ earlier. 

We were interested in fluctuations induced by Gaussian noise. The noise could be additive, in which case the random force is independent of the dynamical variables, or multiplicative, where the random force depends on the instantaneous values of the dynamical variables. The typical intensity of the noise $D$ was the small parameter of the theory. In the analysis of the extinction problem the noise has to be multiplicative, with its strength going to zero at $q_1=0$, to prevent re-emergence of the population once it has gone extinct 

Using the smallness of $D$, we showed that small delay does not
  qualitatively change  small-amplitude fluctuations
about the stable states. We have obtained the spectrum of small-amplitude
fluctuations.

The analysis of large rare fluctuations was reduced to a variational
problem. The solution of this problem gives the logarithm of the probability
distribution on the tail and also the exponents of the rates of switching and
extinction. The rates are described by an activation-type law, with their
logarithm being proportional to the reciprocal noise intensity  $D^{-1}$. The
solution of the variational problem can be associated with the activation
energy of the corresponding transition (switching or extinction). The extreme
trajectories of the variational problem give the most probable paths followed
by a fluctuation to a given state or in switching between the stable states or
in extinction. An important part of the formulation is the boundary conditions
for the extreme trajectories. We found these conditions for systems with delay
and, in particular, found a significant difference in the form of these
conditions for the problems of switching and extinction.

In the presence of delay the variational equations for the extreme
trajectories are acausal: they contain the past and future values of the dynamical variables, which are time-shifted by $\tau$.  This is somewhat reminiscent of the situation with variational
trajectories in the instanton problems for tunneling with dissipation,
cf.~\cite{Caldeira1983}. However, in our case the trajectories go in real
time. They are accessible to direct observation in the experiment, as in
systems with no delay, cf.~\cite{Hales2000,Chan2008}. Numerical evaluation of
the extreme trajectories based on the acausal equations of motion requires
special tools, since shooting methods generally do not work. An approach to
the problem which does not rely on the shooting method was proposed in
Ref.~\onlinecite{Lindley2013}. 

For small delay $\tau$ compared to the relaxation time of the system, the
delay-induced corrections to the activation energies are linear in
$\tau$. This is strongly different from the result for inertial systems with delayed friction force \cite{Dykman2012a}, where the correction to the switching activation energy studied  there was found to be quadratic in small delay. In our system, the corrections exponentially strongly affect the rates of switching
and extinction, since they are in the exponents of the expressions
for the rates and are multiplied by a large factor $D^{-1}$. 

Our results also show that, depending
on the form of the equations of motion, delay can increase or decrease the
corresponding rates. We tested this conclusion using a simple nonlinear model. We found that, in this model, the results of the perturbation theory in $\tau$ agree in a broad parameter range with the results of the direct variational method that we employed. 

We studied the switching and extinction rates in the important parameter range where the system is close to the saddle-node or transcritical bifurcation points.
Because the system slows in the vicinity of equilibria near bifurcations in this range, it is sufficient to look for
the linear in $\tau$ corrections to the activation energies.  We found that
both the leading term in the activation energies and the delay-induced
corrections scale as powers of the distance to the bifurcation point. The
exponents can be the same, or the correction can decrease faster than the
leading-order term as the system approaches the bifurcation point. The
exponents of the leading-order terms and of the corrections are different in the
problems of switching and extinction.

We carried out careful numerical simulations to test the theory. A potential
pitfall in such simulations is that one has to make sure that the system
indeed has switched to another state or indeed has gone extinct. In the
presence of delay, it means that one has to check that the delayed force will
not pull the system back into the domain of attraction of the initially
occupied state. The results of the simulations for the nonlinear models that we employed
are in excellent quantitative agreement with the theory, with no adjustable parameters. This refers to
both the activation dependence of the rate of the transitions on the noise
intensity $D$ and to the dependence of the effective activation energy on
$\tau$. In the studied models, this dependence appeared to be linear in a
comparatively broad range. However, we found that the range where simulations and the perturbative analytical predictions coincide is narrower where the delay is in the term that repels the system from the saddle point and thus drives it back to the attractor even if the system is already behind the saddle point(in the problem of switching) or has reached the extinction state (in the extinction problem).

\section*{Acknowledgments}
IBS gratefully acknowledges support from the Office of Naval Research (N0001414WX00023) and NRL 6.1 Base program (N0001414WX20610). L.B. was supported by the National Science Foundation under CMMI-1233397 and DMS-0959461. MID is supported by US Army
Research Office (W911NF-12-1-0235) and US Defense Advanced Research Agency
(FA8650-13-1-7301). This material is based upon work while LB was serving at
the National Science Foundation. Any opinion, findings, and conclusions or
recommendations expressed in this material are those of the authors and do not
necessarily reflect the views of the National Science Foundation, the ARO, and DARPA.
\bibliographystyle{apsrev4-1}
%

\end{document}